\documentclass[manuscript]{aastex}

\usepackage[colorlinks,
            linkcolor=red,
            anchorcolor=blue,
            citecolor=blue
            ]{hyperref}%,graphicx,times}

\slugcomment{}

\shorttitle{Material supply and magnetic configuration of a filament}
\shortauthors{Zou, P. et al.}

\begin{document}

\title{Material supply and magnetic configuration of an active region
 filament}

\author{P. Zou\altaffilmark{1,2,3}, C. Fang\altaffilmark{1,2,3},
 P. F. Chen\altaffilmark{1,2,3}, K. Yang\altaffilmark{1,2,3}, Q. Hao\altaffilmark{1,2,3}, \& Wenda Cao\altaffilmark{4}}

\email{fangc@nju.edu.cn}

\altaffiltext{1}{School of Astronomy \& Space Science, Nanjing University, Nanjing 210023, China}
\altaffiltext{2}{Key Lab of Modern Astronomy \& Astrophysics (Nanjing University),
                Ministry of Education, Nanjing 210023, China}
\altaffiltext{3}{Collaborative Innovation Center of Modern Astronomy and Space Exploration,
                Nanjing 210023, China}
\altaffiltext{4}{Big Bear Solar Observatory, New Jersey Institute of Technology,
             ~40386 North Shore Lane, Big Bear City, CA 92314, USA}

\begin{abstract}
It is important to study the fine structures of solar filaments with
high-resolution observations since it can help us understand the magnetic and
thermal structures of the filaments and their dynamics. In this paper, we study
a newly-formed filament located inside the active region NOAA 11762, which was
observed by the 1.6 m New Solar Telescope (NST) at Big Bear Solar Observatory
(BBSO) from 16:40:19 UT to 17:07:58 UT on 2013 June 5. As revealed by the
H$\alpha$ filtergrams, cool material is seen to be injected into the filament
spine with a speed of 5--10 km s$^{-1}$. At the source of the injection,
brightenings are identified in the chromosphere, which is accompanied by
magnetic cancellation in the photosphere, implying the importance of magnetic
reconnection in replenishing the filament with plasmas from the lower
atmosphere. Counter-streamings are detected near one endpoint of the filament,
with the plane-of-the-sky speed being 7--9 km s$^{-1}$ in the H$\alpha$
red-wing filtergrams and 9--25 km s$^{-1}$ in the blue-wing filtergrams. The
observations are indicative of that this active region filament is supported
by a sheared arcade without magnetic dips, and the counter-streamings are due
to unidirectional flows with alternative directions, rather than due to the
longitudinal oscillations of filament threads as in many other filaments.
\end{abstract}


\section{INTRODUCTION}

Solar filaments, called prominences when seen above the solar limb, look
like cool and dense clouds suspended in the solar corona.  They are always
located near the polarity inversion lines (PILs), thus can be used to trace the
large-scale pattern of weak background magnetic field \citep{low1982, min1998}.
The long-lived quiescent filaments can also be used to trace the solar
differential rotation \citep{gigo13}. Once they erupt, solar filaments are
often associated with solar flares \citep[e.g.,][]{mar72,sch82} and/or coronal
mass ejections \citep[CMEs,][]{stc1991, chen11}. Many authors have analyzed
their formation, oscillation, eruption, their interaction with other
activities, and the long-term statistical properties \citep{moo1992, man2008,
sch2013, hao15}.

One important issue in the filament research is their formation mechanisms,
which have been studied for decades \citep[e.g.,][]{smi1977}. Generally
speaking, there are three possible mechanisms accounting for the filament
formation \citep{mack10}. The first and also the most popular one is the
evaporation-condensation model. This model has been
extensively studied by many authors using numerical simulations
\citep{ant1999, kar2001, kar2005, kar2006, kar2008,
xia2012, xia2014, zhou14}. The merit of this model is that it can naturally
explain the sudden appearance of the dense cool plasma suspended in the
tenuous hot corona \citep{ber2012, liu12}. The second one is the
levitation model, which was initially proposed by \cite{rust1994}. However,
so far there are only a few observations which can fit into this model
\citep{lit2005, oka2008, yel2012}. For example, \citet{lit1997} observed a
filament being formed after the emergence of a $\delta$-sunspot and suggested
that the cool material is dragged up into the corona through the levitation
process. Instead of being levitated as a whole, it was proposed that
the chromospheric plasmas might be lifted up through tornadoes, which are the
feet of the filaments \citep{li12,su12}. The third one is the injection model,
which suggests that cool material is ejected from the chromosphere into the
corona via magnetic reconnection \citep{wang1999}. Some observations revealed
strong flows moving into active-region filaments, which tend to
support this model \citep{chae2003, sch2004}. However, it seems that the
larger and higher quiescent filaments are not formed by this mechanism
\citep{kar2015}. This is probably because that the injection speed of the
cool plasma is not high enough for the materials to reach the magnetic dips
of a quiescent filament. In this sense, the velocity measurements of the cool
material around filaments are important for understanding the feasibility
of the injection mechanism.

One interesting feature of the filament dynamics is the
counter-streamings \citep{zir1998}. A natural interpretation is that the
counter-streamings are caused by longitudinal oscillations of filament
threads \citep{kar2006, xia11}. In this case, a magnetic dip is needed, which was
originally proposed to support the apparently static filament in the
low-resolution images. The magnetic dips may exist in a flux rope \citep{kr74}
or a sheared arcade \citep{ks57}. However, with numerical simulations,
\citet{kar2001} proposed that a magnetic dip is not necessary for supporting a
filament, where cool plasmas repeatedly appear and disappear near magnetic
PILs. In this case, longitudinal oscillations are not sustainable, so the
counter-streamings, which were found to be universal in both quiescent and
active region filaments, may be due to large-scale unidirectional flows,
which have opposite moving directions in neighboring flux tubes, as
proposed by \citet{chen2014}. However, there is still lack of evidence that
counter-streamings are in the form of unidirectional flows with alternative
directions. Since a quiescent filament
is typically supported by a magnetic flux rope where magnetic dips are
always present, it is expected that the unidirectional
flow-based counter-streamings can be seen only in active region filaments.

As the spatial resolution increases, observations are providing more
and more detailed structures and dynamics of solar filaments, which are helpful
to unveil the formation mechanism and the material circulation in filaments.
In this paper, we use observations from the current largest solar telescope,
1.6 m New Solar Telescope \citep[NST,][]{cao2010,goo2012}, at the Big Bear Solar
Observatory (BBSO) to analyze the material flows in a newly-formed filament
by tracing its fibrils. It is expected to give an insight into the
questions about where the filament material comes from and how the
counter-streamings are formed in the filament. The observations are described
in Section~\ref{observation}. The results are given in Section~\ref{results}.
The discussions and the summary are presented in Section~\ref{discussion}.

\section{OBSERVATIONS}
\label{observation}

\begin{figure}
\centering
\includegraphics[width=14cm,trim = 0 0 0 0,clip = true]{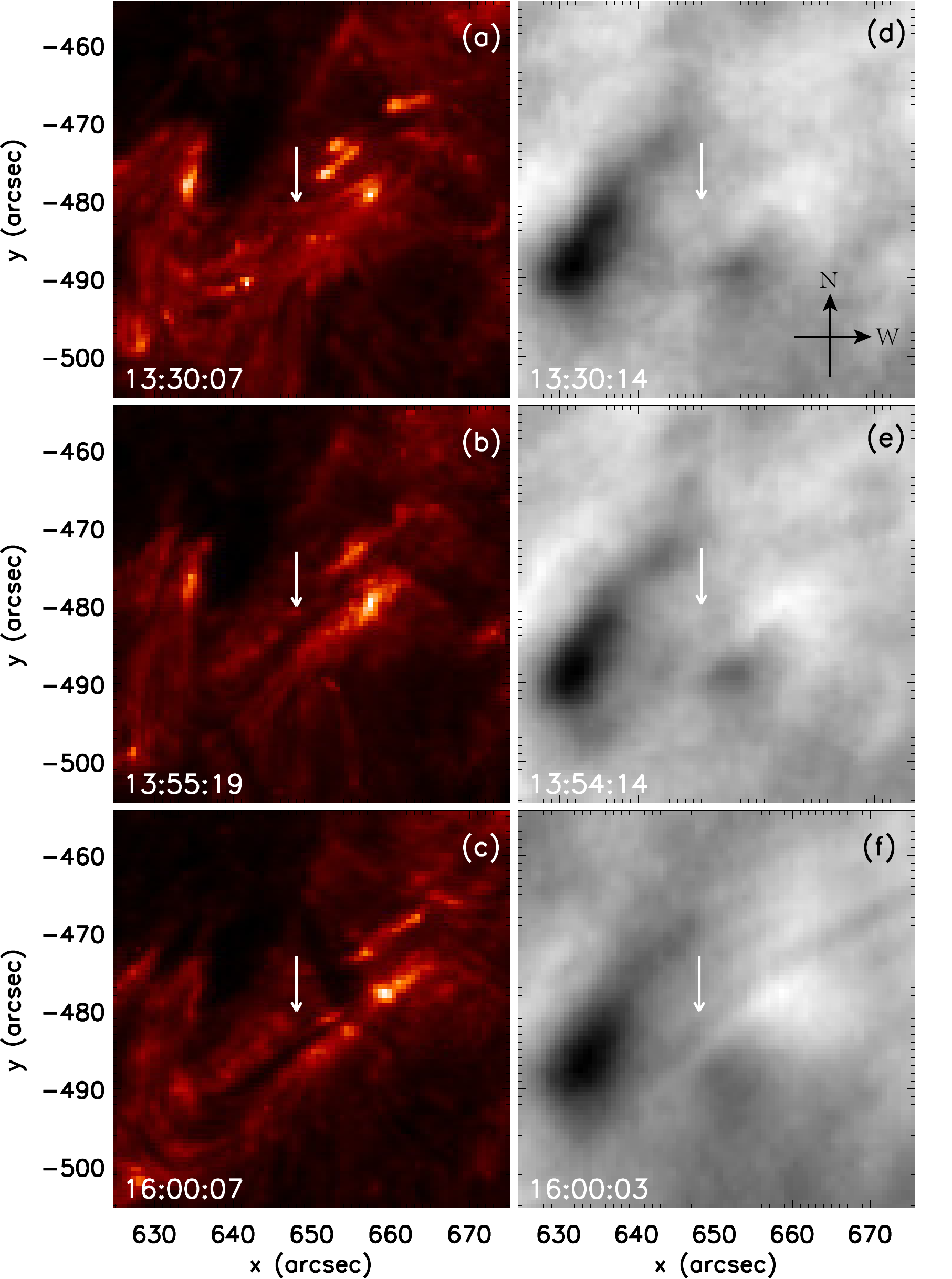}
\caption{Locations and temporal evolution of the filament in 304 \ \AA\ and H$\alpha$ images. In panel (a) the filament is unseen in the 304 \AA\ image. In panel (b) the filament is firstly seen and then can be clearly seen in panel (c). In panel (d) the filament is still unseen in the full-disc H$\alpha$ image, and some fragments can be seen in panel (e). However, it can be obviously seen in panel (f). }
\label{fig1}
\end{figure}

The newly-formed filament was located in the active region NOAA 11762 (S28W62)
on 2013 June 5. Its formation is well recorded by the Atmospheric Imaging
Assembly \citep[AIA,][]{leme12} in EUV and the Global Oscillation Network Group
({\it GONG}) in H$\alpha$ \citep{harv11}. AIA is on board the \emph{Solar
Dynamics Observatory} ({\it SDO}). The AIA telescope has a pixel size of
$0\farcs 6$ with a time cadence of 12 s. As indicated by the arrow in Figure
\ref{fig1}(b), the filament is first seen in AIA 304 \AA\ at 13:55 UT. At that
time, nothing is visible in H$\alpha$, as shown by Figure \ref{fig1}(e). The
filament spine becomes visible in H$\alpha$ after 16:00 UT, as indicated by
Figure \ref{fig1}(f).

The filament was also observed by the NST at BBSO from 16:40:19 UT to 17:07:58
UT with a much higher resolution. The field of view (FOV) is centered at
(647\arcsec, -482\arcsec) in the heliocentric coordinates. NST consists of a
suite of instruments. Its Broadband Filter Imager (BFI) uses a TiO filter
(7057 \AA) to obtain the white-light images with a bandpass of 10 \AA.
It has a time cadence of 15 s. The diffraction-limited resolution
is $0 \farcs 11$ and the pixel size is $0 \farcs 034$. The filtergrams
at H$\alpha$ line center, H$\alpha \pm$0.2 \AA, $\pm$0.6 \AA, and
$\pm$1.0 \AA\ are acquired by the Visible Image Spectrometer (VIS).
The VIS uses a Fabry-P\'erot etalon with the bandpass of 0.07 \AA.
Its FOV is $70\arcsec \times 70\arcsec$. The diffraction limited resolution
is $0 \farcs 1$ and the pixel size is $0 \farcs 03$.

To understand the magnetic environment of the filament formation and its flows,
we use the magnetograms observed by the Helioseismic and Magnetic Imager
\citep[HMI,][]{sch2012,schou2012} on board the {\it SDO}. In particular, one
special product of the HMI data, i.e., the Space-weather HMI Active Region
Patches \citep[SHARPs,][]{bob2014}, is used, where the minimum energy method is
adopted to resolve the 180$^{\circ}$ ambiguity \citep{met1994,met2006}. The
Lambert method is applied to modify the coordinate system \citep{bob2014}, and
the projection effects are corrected with the method used by \cite{gar1990}.

\section{RESULTS}\label{results}

\begin{figure}
\includegraphics[width=17cm,trim = 35 0 0 0,clip = true]{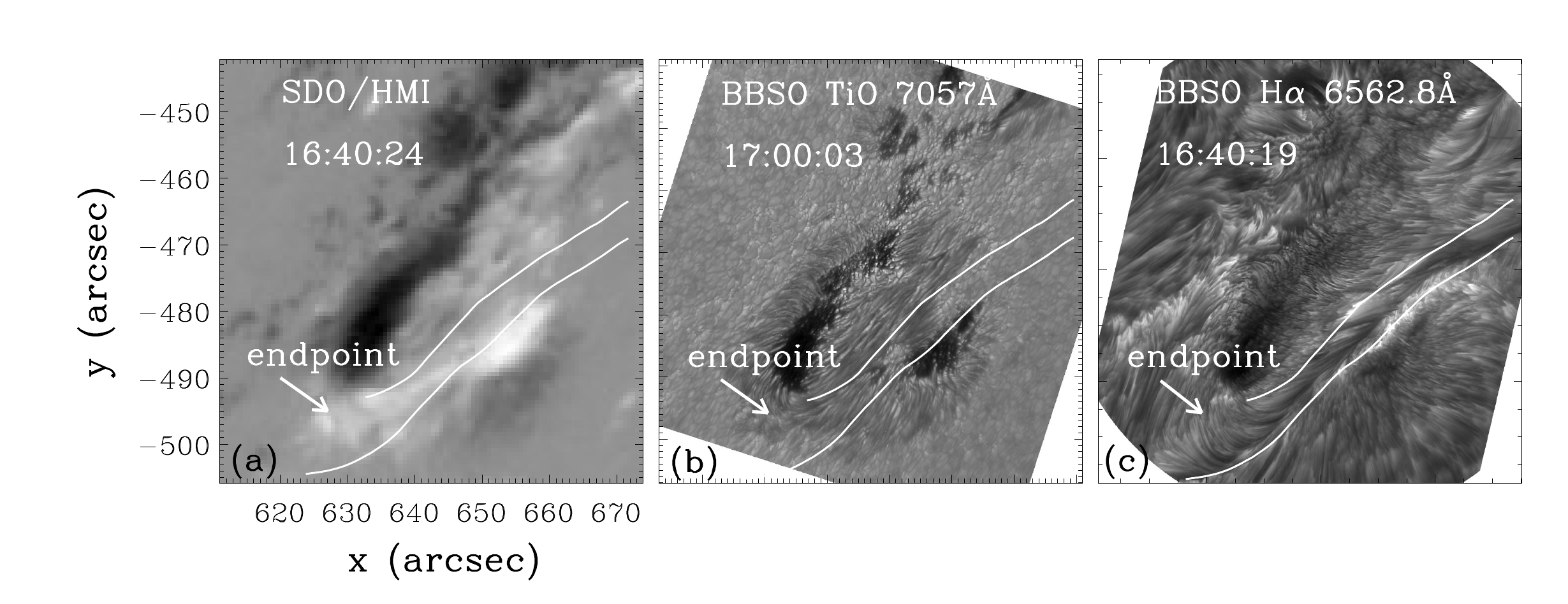}
\caption{Co-aligned the images of magnetogram (a), TiO (b), and H$\alpha$ line-center (c). The white lines in each panel sketch the profile of the filament. The filament endpoint is also indicated in each panel.}
\label{fig2}
\end{figure}

Using the AIA 4500 \AA\ and NST TiO images, we co-align the images in TiO
and H$\alpha$ with the magnetograms observed by {\it SDO}/HMI. Figure
\ref{fig2} displays the co-aligned magnetogram ({\it left panel}), white-light
image in TiO ({\it middle panel}), and the H$\alpha$ center filtergram
({\it right panel}). The rough location of the filament is bracketed
by two white lines. It can be seen that the filament is located near the
magnetic PIL between two sunspots, with the southern endpoint near the
sunspots and the other endpoint outside the field of view (FOV). The
filament has an ``S"-shaped structure, a typical one with positive
helicity mainly in the southern hemisphere. As seen from the right panel of
Figure \ref{fig2}, this filament has many right-bearing barbs, and the
threads are right bearing as well.

\subsection{Plasma motion and bright patches}

In the H$\alpha$ line-wing images, we can identify many moving fibrils on
both sides of the filament spine. In Figure \ref{fig3}(a), we select
three small regions, which are displayed in panels (b)--(d), respectively.
Several moving fibrils can be clearly seen in the H$\alpha$ line-wing
filtergrams, as marked by the dashed lines in Figures \ref{fig3}(b)--(d). In all
these fibrils, materials are seen to be injected toward the filament spine. One
strange feature is that the fibrils on the two sides of the filament spine are
far from being parallel. They even have opposite senses of bearing, according
to the definition by \citet{mart94}, i.e., the fibrils on the east side are
right-bearing, whereas those on the west side are left-bearing.

After examining Figure \ref{fig3} and the associated movie, we found that
the material injections along the fibrils generally originate from bright
regions. In order to illustrate it, we plot the H$\alpha$ center
filtergram in Figure \ref{fig4}(a), where the areas with intensity enhancement
are highlighted with yellow contours as done in Figure \ref{fig3}(a). Inside
the red rectangular box, the brightness is integrated over all the pixels with
relative intensity higher than 2 (the averaged intensity of the background is
set to be unity), and the temporal evolution of the integrated intensity is
shown in panel (c) as the solid line. It is seen that the integrated
brightness changes drastically with time. At the same time, the underlying
photospheric magnetic field also changes. In order to check whether
they are related to each other, we select the blue rectangular box in
Figure \ref{fig4}(b) to calculate the total negative magnetic flux inside the
box. The blue box is shifted from the red one after considering the
projection effects and assuming that the H$\alpha$ formation layer is about
2000 km above the photosphere where the magnetic field is measured.  The
variation of the negative magnetic flux inside the blue box is displayed in
panel (c) as the dashed line. It can be seen in Figure \ref{fig4}(c)
that almost each episode of H$\alpha$ brightening is preceded by magnetic flux
cancelation. Using the cross-correlation method, the lightcurve and the
negative magnetic flux evolution are found to be negatively correlated, with
the maximum correlation coefficient being -0.76 when the time delay between the
lightcurve and the magnetic flux is set to be 320 s. That is to say, each time
when magnetic cancelation happens, H$\alpha$ brightness increases about 320 s
later. Thus, we conjecture that the brightening of the filament channel is
possibly caused by magnetic field cancelation. It is noted that since
the resolution of our H$\alpha$ observations is about twenty times higher than
that of the HMI magnetogram, it is nearly impossible to compare the detailed
structure in H$\alpha$ filtergrams with that in the magnetogram. To establish a
more quantitative relation between the magnetic cancelation and the
chromospheric heating, future magnetograms with higher resolutions are needed.

Now we select one typical fibril to analyze its dynamics. This fibril,
labeled fibril 1, is located on the west side of the filament spine, as shown
in Figure \ref{fig3}(b). A slice is placed along fibril 1 in Figure 
\ref{fig5}(a). The time-distance diagram of the H$\alpha$ intensity along this
slice is displayed in Figure \ref{fig5}(b). Since the flow is more evident in
the H$\alpha$ line wing and the chromospheric brightening is more visible in
the H$\alpha$ line center, H$\alpha+0.6$ \AA\ intensity image is used for the 
top half distance, whereas H$\alpha$ center intensity image is used for the 
bottom half distance. It is seen that several moving fibrils exist in the top 
half distance and several H$\alpha$ brightenings exist in the bottom half 
distance. It is indicative of that the ejection and the brightening 
should be physically connected. We tentatively draw three green arrows in order
to match H$\alpha$ fibrils with the corresponding brightenings. Besides, there are another two moving fibrils at $t=10$ s and 150 s, which should be related 
to chromospheric heating before and after our observations, respectively.

It can be seen in Figure \ref{fig5} that the material injections are episodic.
Each fibril lasts for about 100--200 s. The injection velocities are
measured to be about 5--10 km s$^{-1}$. \citet{chae2003} proposed that
reconnection-driven jets in the low atmosphere might be able to supply mass
for the formation of a solar filament. The fibrils revealed by our Figure
\ref{fig5} tend to support such a scenario.

\subsection{Counter-streamings and magnetic configuration}

In contrast to the main spine where threads are highly sheared and overlapping
with each other, the threads near the southern endpoint of the filament are
more clearly separated, which offers a good chance to study their dynamics. The
H$\alpha$ movie indicates that these threads are filled with moving materials.
In order to investigate their dynamics, we choose two slices and
display their time-distance plots in different wavelengths in Figure
\ref{fig6}. Both slices start from the spine (bottom right) to the endpoint
(upper left) following the threads near the filament endpoint. It can be seen
that along the red slice, repetitive cool plasma flows move to the south. On
the contrary, the cool plasma in the yellow slice
moves to the north. We further select several other slices covering different
threads, and calculate the moving velocities based on time-distance
diagrams. The resulting velocities are plotted as arrows superimposed on the
H$\alpha$ filtergram in Figure \ref{fig7}. The length of each arrow represents
the amplitude of the velocity of the plasma flow, and the color indicates which
off-line filtergram can best show the moving plasma. The velocity
pattern in Figure \ref{fig7} is strongly reminiscent of the counter-streamings
discovered by \cite{zir1998}. The velocities of the southward flows are about
7--9 km s$^{-1}$ and those of the northward flows are about 9--25 km s$^{-1}$
(see Figure \ref{fig7}).

\section{DISCUSSIONS}\label{discussion}

\subsection{Mass supply and counter-streamings}

The formation mechanism of solar filaments is an important issue in
filament research \citep{pare14}. Although an individual author might favor
one mechanism, it seems that filaments may be formed in different ways.
For those filaments that condense in situ in the corona \citep{liu12}, the
thermal nonequilibrium model works well \citep{kar2006, xia11, xia2012}, which
can qualitatively explain the in-situ brightening successively from
high-temperature extreme ultraviolet (EUV) lines to low-temperature EUV lines,
and all the way to the H$\alpha$ line \citep{chen14a}. According to this
model, chromospheric plasma is evaporated into the corona due to localized
heating, forming hot and dense coronal loops. When the density reaches a
threshold, thermal instability is triggered, and the hot plasma cools
drastically to form cool plasma suspended in the corona. On the other
hand, there are also accumulative examples of filament formation via injection,
i.e., cool chromospheric plasmas are injected into the corona as a result of
magnetic reconnection \citep{chae2003,sch2004}. In this case, the typical
features include (1) low atmosphere brightening can be seen in H$\alpha$ or
even EUV; (2) cold plasma is driven from the chromosphere to the corona so as
to supply materials for the filament. Our observations presented in this paper
are consistent with this model.

We found that on the both sides of the filament channel, there are sustaining
brightening visible in H$\alpha$ line-center images. From time to time,
materials are injected upward from these bright patches to replenish the
filament spine with a velocity of about 5--10 km s$^{-1}$. In particular, near
the south endpoint of the filament, the injected materials form the threads of
the filament directly.  The injections are intermittent, with an interval from
tens of seconds to several minutes. Each injection lasts for about 100--200 s,
preceded by an enhanced brightening in H$\alpha$ at the source site, indicating
localized heating in the chromosphere. Note that the ejected materials
move upward along the magnetic field and should be observed as blue-shift if 
the ejection is toward the observer. However, the flows are observed to be
red-shifted. The reason is that this active region is located in the southwest
quadrant of the solar disk, and the magnetic field lines are bent away from
the observer (as indicated by Figure \ref{fig8}b). As a result, the 
upward-moving materials are observed to be red-shifted. Compared with the
photospheric
magnetograms, it is found that the brightening is almost always associated with
magnetic cancelation, implying that magnetic reconnection is happening
in the low atmosphere \citep{wang09}. A rough estimation indicates that
the injection velocity is too small to lift the material into the corona if
the plasma experiences a ballistic motion with the gravity acting on it only.
The fact that the cool materials keep moving without significant deceleration
implies that there must be additional forces acting on the flows. Since the
cold plasma flows move along the magnetic field, the Lorentz force does not
help. A possible
source of the additional forces could be the higher gas pressure in the
source region. Through 3D magnetohydrodynamics (MHD) numerical simulations,
\citet{jian11} found that whereas the reconnection jet is accelerated initially
by magnetic tension force, its later evolution is dominated by the gas pressure
gradient. In our observations, chromospheric brightenings appear in the source
site of the injection. Its enhanced gas pressure might provide an additional
force to sustain the moving materials against gravity. The enhanced pressure
may even trigger a shock wave, which can effectively push the fibrils upward
\citep{ryut08}.

As the characteristic dynamics of a filament in the quiescent state,
counter-streamings are generally explained in terms of longitudinal
oscillations of filament threads \citep{lin2003}. It is generally believed
that there exist dips in the magnetic field of the filaments, which enable the
gravity to be a restoring force and facilitate the longitudinal oscillations
\citep{luna12, zhan12}. However, inspired by the observations showing
that filaments are always in a dynamic state, \citet{kar2001} pointed out that
magnetic dips might not be a necessary condition for filaments. They proposed
that, driven by asymmetric heating at the two footpoints of a magnetic tube, 
chromospheric plasma is evaporated into the corona. The condensed hot gas
cools down to form a filament thread. After staying in the corona for some
time, the filament thread is pushed by the gas pressure imbalance between the
two footpoints of the magnetic tube, draining down toward the footpoint with
weaker heating. In this model, an ensemble of threads like this form the
counter-streamings. The observations displayed by \citet{liu12} seem to support
such a dynamic picture, i.e., filament materials are circulated via
chromospheric evaporation, coronal condensation, and mass drainage. In this
paper, we presented another scenario: cold chromospheric materials are
intermittently injected into the corona from one footpoint of a magnetic tube,
which run through the flux tube in the corona and then fall down toward the
other footpoint. A random distribution of the injection sources results in
alternative flows from the positive to the negative polarities and the other
way around, leading to counter-streaming pattern displayed in Figure
\ref{fig7}. In this scenario, no chromospheric evaporation is required.

\subsection{Magnetic Configuration}

Two types of magnetic configurations have been proposed for solar filaments,
i.e., a flux rope \citep{kr74} and a sheared arcade \citep{ks57}.
Traditionally, the difference between these two configurations
can be distinguished by measuring the magnetic component of filament threads
perpendicular to the magnetic PIL. That is, the flux rope corresponds to
a magnetic field with inverse polarity, whereas the sheared arcade corresponds
to a magnetic field with normal polarity. Recently, \citet{chen2014} proposed
an indirect method to infer the magnetic configuration of a filament: a filament
with left-bearing barbs and positive helicity (or with right-bearing barbs and
negative helicity) is supported by a flux rope, whereas a filament with
right-bearing barbs and positive helicity (or with left-bearing barbs and
negative helicity) is supported by a sheared arcade\footnote{It should
be noted here that \citet{mart08} claimed that right-bearing barbs correspond
to dextral filaments and left-bearing barbs correspond to sinistral filaments.
As pointed out by \citet{chen2014}, this one-to-one correspondence is valid
only for the inverse-polarity filaments, i.e., those supported by a magnetic
flux rope. For the normal-polarity filaments, the correspondence should be
exactly opposite.}. With the method proposed by \citet{chen2014},
\citet{ouya15} and \citet{hao16} identified the magnetic configurations of
several filaments, with some being hosted by a flux rope, and others by a
sheared arcade. When applying this model, we have to measure the sign of
helicity of the filament beforehand.

In order to calculate the sign of helicity, the vector magnetic field data
SHARPs, with the 180$^{\circ}$ ambiguity resolved and projection effects
corrected, are used to calculate the twist parameter $\alpha=(\nabla \times
{\bf {B}})_z/B_z$, where ${\bf B}$ is the vector magnetic field and $B_z$ is
the line-of-sight component of $\bf B$. It was found that $\alpha$ in the
filament channel we observed is positive, which is the preferential sign of
helicity in the southern hemisphere. Since the barbs of our filament are
right-bearing as indicated by Figure \ref{fig4}, the filament should be
supported by a sheared arcade based on the indirect method proposed by
\citet{chen2014}.

Coronal magnetic extrapolation, albeit an ill-posed problem \citep{low15},
provides another method to examine the host magnetic field supporting a
filament. In order to perform the non-linear force-free field (NLFFF)
extrapolation, we apply the optimization method \citep{whe2000,wie2004}
with the SHARPs vector magnetogram. To remove the net force and torque on
the boundary before extrapolation, data pre-processing is conducted with the
method mentioned by \citet{wie2006}. The NLFFF distribution in the
local Cartesian coordinates is shown in the top panels (a) and (b) of Figure
\ref{fig8}, where magnetic field lines are superposed on the H$\alpha$-0.6
\AA\ filtergram. The field lines are selected near the filament spine. It can
be seen that the magnetic configuration near the filament spine consists of a
strongly-sheared arcade as the core field ({\it cyan}), which is surrounded by
less-sheared envelope field ({\it yellow}). The magnetic configuration around
the south endpoint consists of some less-sheared arcades ({\it green}). The
magnetic configuration projected onto the plane-of-the-sky is displayed in
panel (b) of Figure \ref{fig8}.

It is noticed that no magnetic flux ropes are present in the extrapolated
nonlinear force-free coronal field. Although the absence of magnetic dips in
the extrapolated coronal field might be due to the limitation of the
extrapolation method, combining the result of the repetitive moving threads
near the southern endpoint of the filament, we tend to believe that the
filament, at least the southern part (the green lines around the south
endpoint in H$\alpha$-0.6 \AA\ filtergram), is supported by a sheared arcade
without magnetic dips.

\begin{figure}
\centering
\includegraphics[width = 15cm,trim = 0 0 0 0,clip = true]{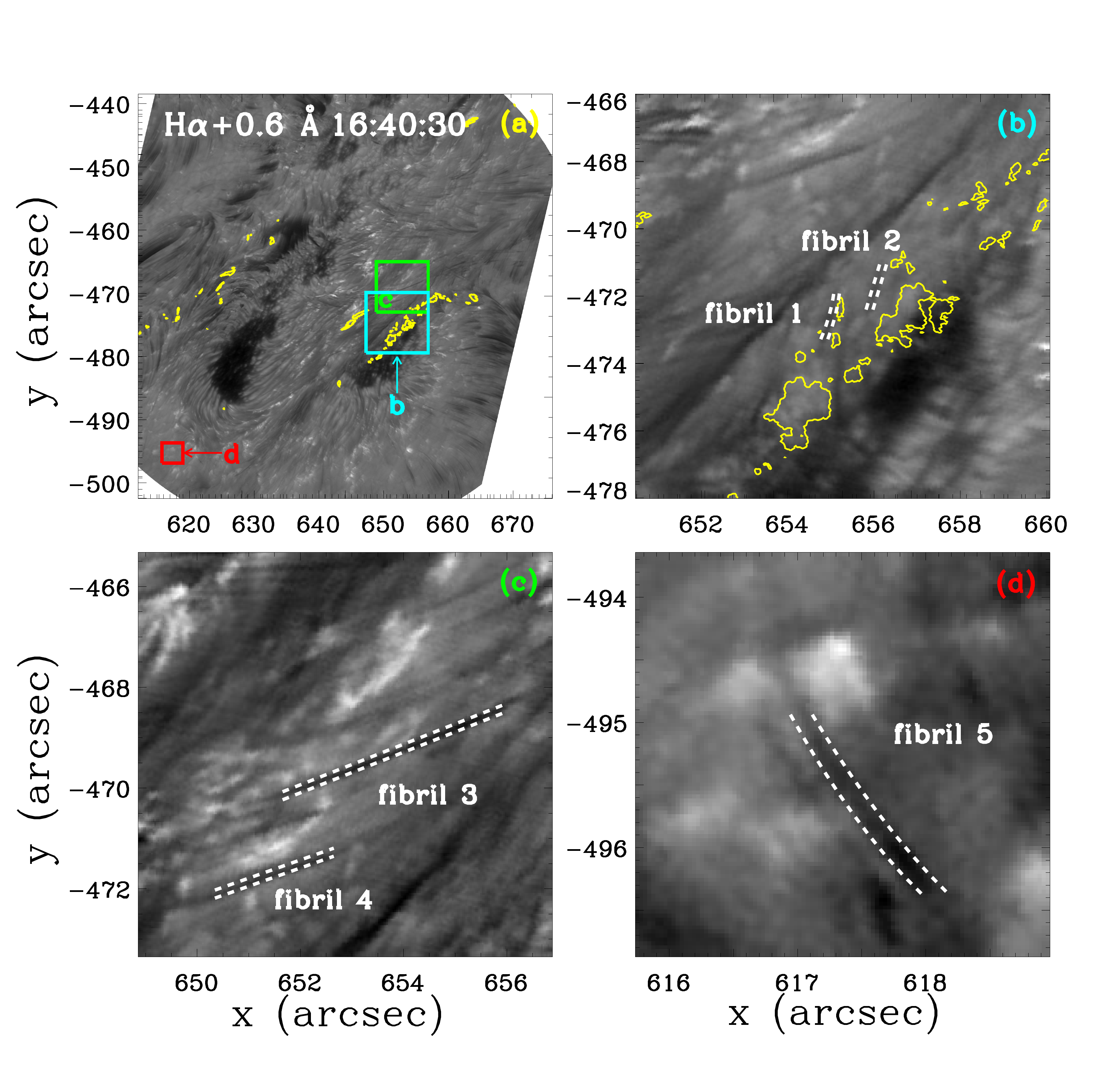}
\caption{Selection of several fibrils near the filament channel. Panel (a) depicts the filament spine and the areas
where we select fibrils (boxes b and c) and the endpoint of the filament (box d). Panels (b), (c), and (d) are
amplified from the boxes in panel (a). The white dashed lines outline the fibrils.}
\label{fig3}
\end{figure}

\begin{figure}
\centering
\includegraphics[width=14cm,trim = 0 0 0 0,clip = true]{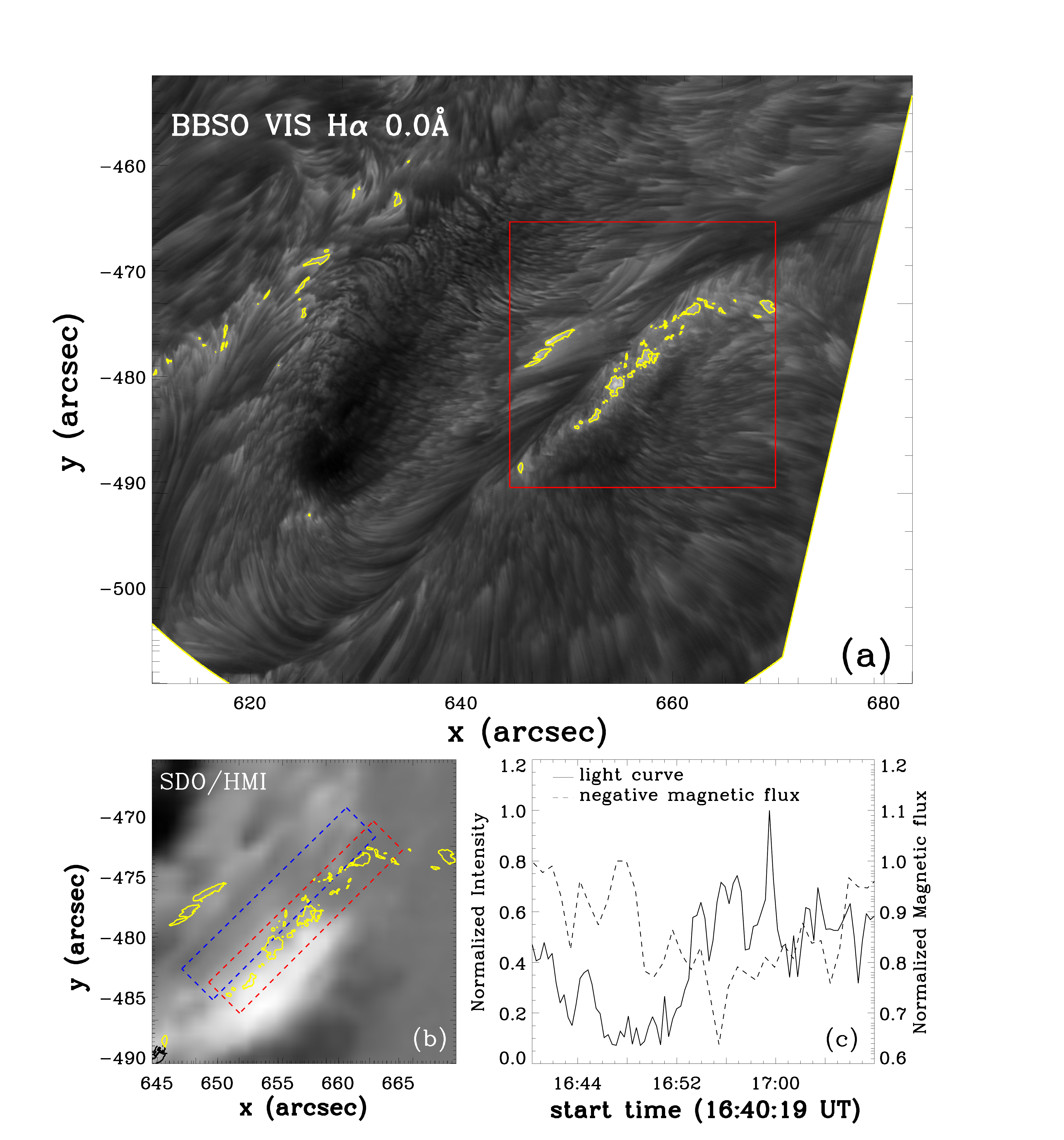}
\caption{H$\alpha$ line-center filtergram (a) and the associated
magnetogram (b). In panel (a), bright patches are highlighted with the
yellow contours, and the red box corresponds to the area of magnetogram we plot
in panel (b). The red rectangular box in panel (b) shows the area we
selected to calculate the lightcurve. A blue rectangular box in panel (b) shows
the corresponding area in the magnetogram, with the project effect corrected.
Panel (c) displays the lightcurve of the bright patches (solid line; integrated
all pixels with relative intensity about twice over that of the background)
and magnetic flux evolution (dashed line; integrated all the negative magnetic
flux), respectively. Both curves are normalized.}
\label{fig4}
\end{figure}

\begin{figure}
\includegraphics[width=17cm,trim = 0 0 0 0,clip = true]{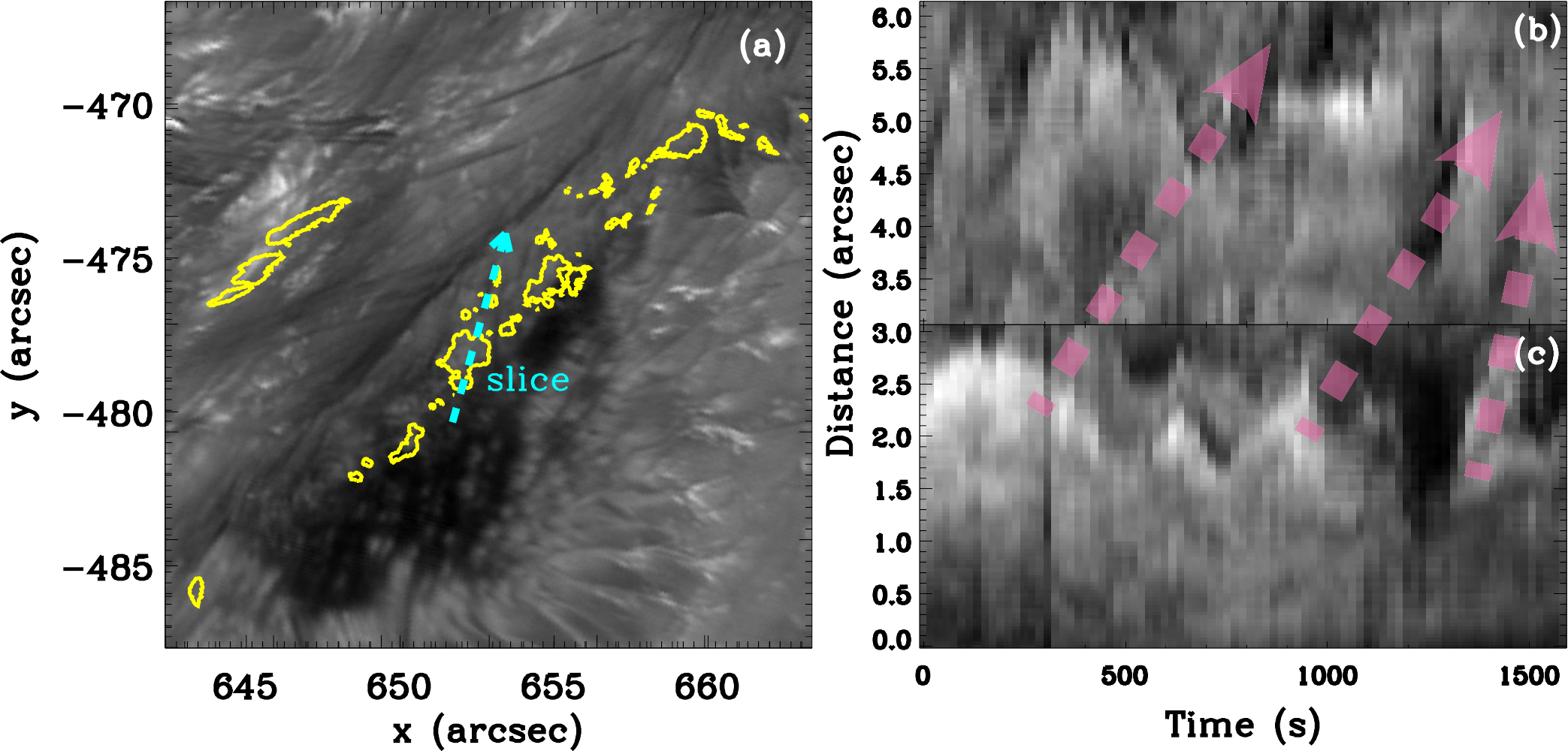}
\caption{Position of the slice along the fibril 1 (panel (a); cyan dashed arrow) and its time-distance plot 
(panels (b) and (c)). Panel (b) is plotted by using the H$\alpha$+0.6 \AA\ filtergram where the jet-like fibril 
is clearly seen. Panel (c) is plotted by using H$\alpha$ center filtergram which gives the information of 
bright patches (yellow contour in panel (a)). Green arrows connect the bright patches and fibrils to show that 
an intensity enhancement in H$\alpha$ center corresponds to a fibril seen in H$\alpha$ red-wing filtergram.}
\label{fig5}
\end{figure}

\begin{figure}
\includegraphics[width=17cm,trim = 0 0 0 0,clip = true]{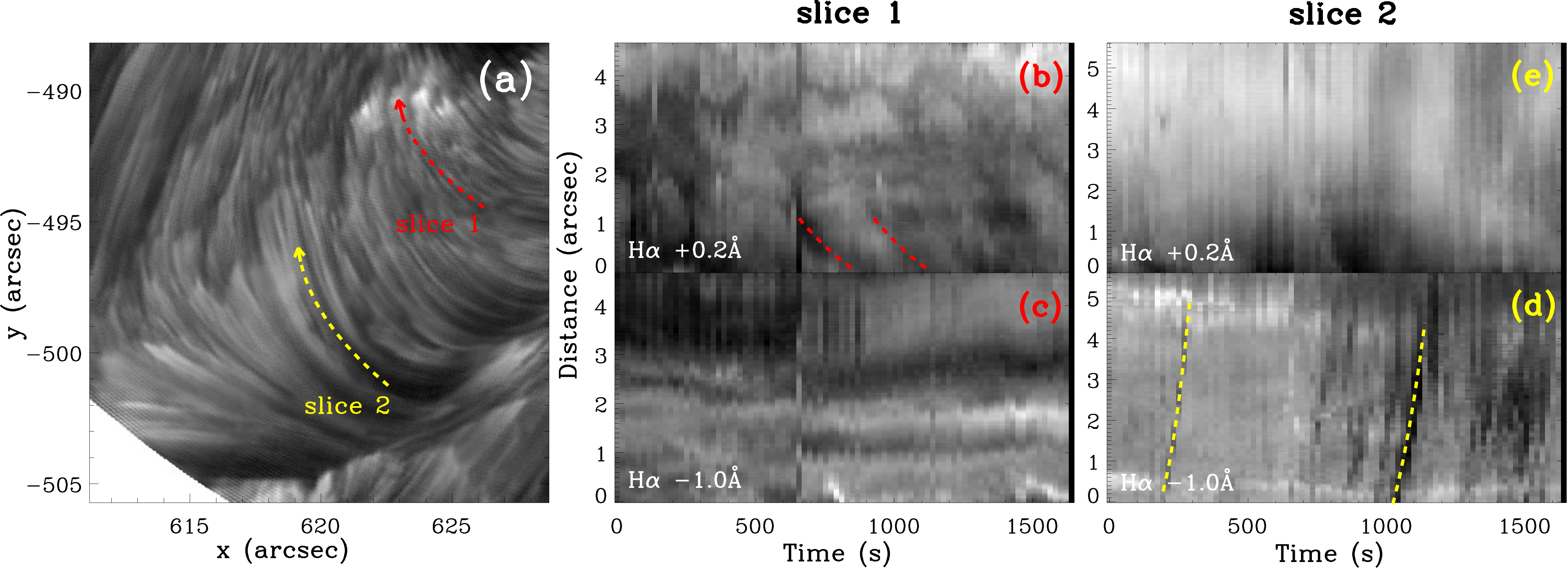}
\caption{Time-distance diagrams of two fibrils near the endpoint of the filament (middle and right column). The left panel depicts the positions of two slices in different wavelengths. The middle column shows time-distance diagrams of slice 1 in red and blue wings and the right column shows the same information of slice 2. We combine these two columns to show the counter-streamings with opposite directions and they also indicate that these fibrils are unidirectional flows.}
\label{fig6}
\end{figure}

\begin{figure}
\plotone{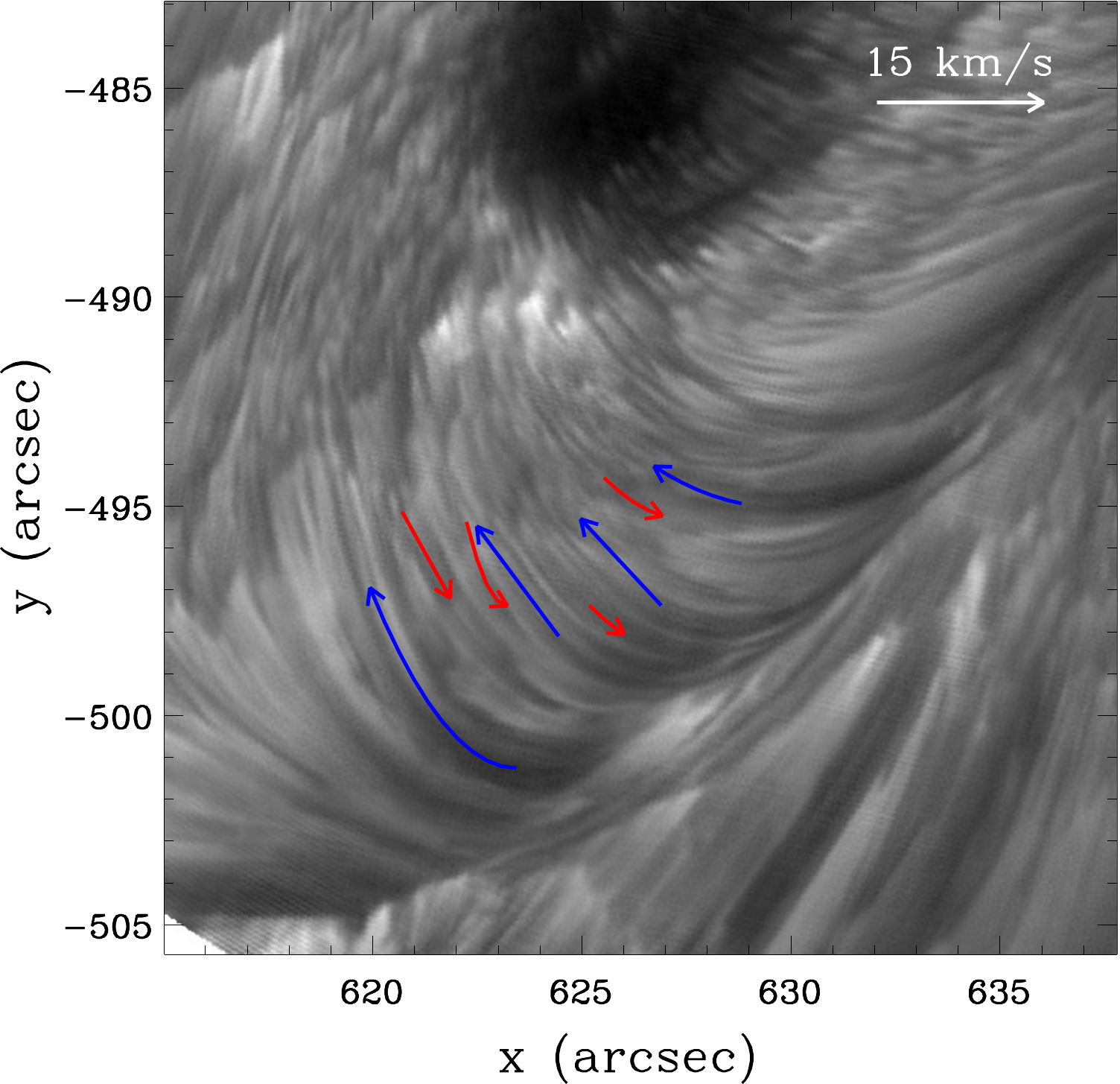}
\caption{Flows near the endpoint of the filament. The red and blue arrows indicate the flow observed in H$\alpha$ red-wing and blue-wing filtergrams, respectively. The length of each arrow is proportional to the local speed.}
\label{fig7}
\end{figure}

\begin{figure}
\includegraphics[width=17cm,trim = 0 50 0 0,clip = true]{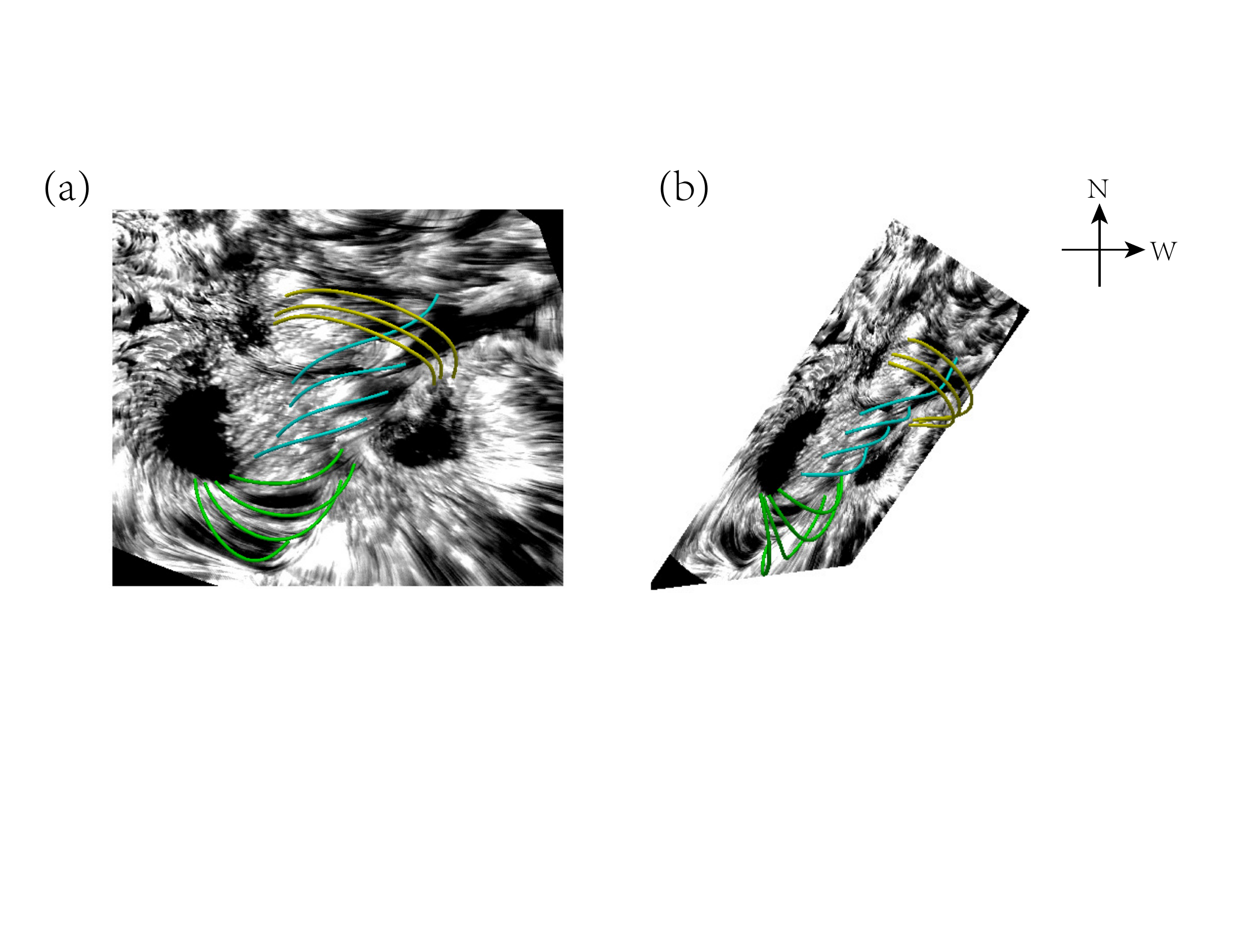}
\caption{Different views of the NLFFF configuration. Panel (a) depicts the top view in the local Cartesian coordinators. Panel (b) shows the magnetic field lines projected onto the plane-of-the-sky. The backgrounds in both panels are the H$\alpha$-0.6 \AA\ filtergram. North is upward.}
\label{fig8}
\end{figure}

Figure \ref{fig3} reveals that the fibrils on the west side of the filament
spine are left-bearing, opposite to the fibrils
on the east side of the filament which are right-bearing. Comparing Figure
\ref{fig3} with Figure \ref{fig8}(b), one can see that the apparent
contradiction might be simply due to the projection effects: the south leg of
the yellow field lines in Figure \ref{fig8}(b), which are oriented toward
northwest as the H$\alpha$ fibrils do in Figure \ref{fig3}(b), are actually
oriented toward northeast in the local coordinates, as illustrated by
Figure \ref{fig8}(a). Therefore, the H$\alpha$ fibrils on both sides of the
filaments are actually right-bearing.

In summary, we analyzed the NST high-resolution observations of a newly
formed active region filament. Based on the observations, it is shown that
cool materials are injected in the form of fibrils from the chromosphere to
replenish the filament suspended in the corona. Each material
injection is preceded by a localized brightening in the chromosphere. It is
caused by magnetic cancelation in the photosphere, implying that magnetic
reconnection plays an important role in transporting chromospheric plasma into
the filament. We also detected counter-streamings near the southern endpoint of
the filament. We argue that the counter-streamings in this active region
filament may be explained by unidirectional flows with alternative directions.
This is different from many of other cases where the counter-streamings are
mainly due to longitudinal oscillations of the filament threads. Nonlinear
force-free field extrapolation further leads us to the conclusion that
this active region filament is supported by a sheared arcade without magnetic
dips. More events are being collected from the observations of NST, New Vacuum
Solar Telescope \citep[NVST;][]{liu2014}, and Optical and Near-infrared Solar
Eruption Tracer \citep[ONSET;][]{fang2013} in order to see how common these
features are among active region filaments.

\acknowledgments
This work was supported by the National Natural Science Foundation of China
(NSFC) under the grant numbers 11533005, 10673004, 10610099, and 11025314, as
well as NKBRSF under grants 2011CB811402 and 2014CB744203. W.C. acknowledges
the support of the US NSF (AGS-0847126 and AGS-1250818) and NASA (NNX13AG14G).
This work was also supported by the project ``The Strategic Priority Research
Progran of the Chinese Academy of Sciences" (XDB09000000).

\clearpage

\clearpage

\end{document}